\documentclass[11pt]{article}
\usepackage{amssymb,amsmath,amsfonts}
\usepackage{graphicx}
\usepackage{graphics}
\usepackage{eepic,epsfig}
\usepackage{verbatim}
\usepackage{color}
\usepackage{hyperref}
\usepackage{cite}

1.60cm \makeatletter \@addtoreset{equation}{section}

\makeatother

\begin{document}
\title{Vacuum bosonic currents induced by a 
compactified cosmic string in dS background}
\author{E. A. F. Bragan\c{c}a$^1$\thanks{E-mail: eduardo.braganca@uemasul.edu.br} ,  E. R. Bezerra de Mello$^2$\thanks
{E-mail: emello@fisica.ufpb.br} \ and \ A. Mohammadi $^3$\thanks{E-mail: azadeh.mohammadi@ufpe.br}\\
\\
\textit{$^{1,3}$Departamento de F\'{i}sica, Universidade Federal de Pernambuco,}\\
\textit{52171-900, Recife-PE, Brazil}\\
\\
\textit{$^{2}$Departamento de F\'{\i}sica, Universidade Federal da Para\'{\i}ba}\\
\textit{58.059-970, Caixa Postal 5.008, Jo\~{a}o Pessoa, PB, Brazil}\vspace{%
0.3cm}}
\maketitle
%
\begin{abstract}
%
In this paper, we investigate the vacuum bosonic currents in the geometry of a
compactified cosmic string in the background of the de Sitter spacetime. The
currents are induced by magnetic fluxes, one running along the cosmic string and another one enclosed by the compact dimension. In order to develop this analysis, we obtain the complete set of normalized bosonic wave-functions obeying a
quasiperiodicity condition. In this context, we calculate the  azimuthal and axial current densities. Due to the quasiperiodicity condition, the quantum number associated with the compactification of the string along its axis becomes discrete, and we use the Abel-Plana summation formula to evaluate the non-vanishing current densities.
We show that these quantities are explicitly decomposed into two contributions:
one corresponds to the geometry of a straight uncompactified cosmic string
and the other is induced by the compactification. We also compare the results with the literature in the case of a massive fermionic field in the same geometry. 
\end{abstract}
\bigskip


\bigskip
%
\section{Introduction}
\label{Int}
%
De Sitter (dS) is a curved spacetime which has been most analyzed in the context of quantum field theory. The main reason is due to its degree of symmetry which allows that many physical problems to be exactly solvable.\footnote{De Sitter space enjoys the same number of degrees of symmetries as the Minkowski one \cite{BD}.}
In addition, the importance of these theoretical analyses increased by the appearance of the inflationary cosmological scenario for
the Universe expansion in its early stages \cite{Linde90}. During an inflationary epoch, quantum fluctuations in the inflaton field introduced inhomogeneities and may affected the transition toward the true vacuum. These fluctuations play important role in the generation of cosmic structures from inflation.

Cosmic strings are linear gravitational topological defects which may have
been created as a consequence of phase transitions in the early Universe and are predicted in the context of the standard gauge field theory of elementary particle physics \cite{VS,hindmarsh,Hyde:2013fia}. Although the observations of anisotropies
in the cosmic microwave background radiation by COBE, WMAP and more recently by the Planck satellite have ruled out cosmic strings as the primary source for primordial density perturbations \cite{Ade:2013xla},
this topological deffect can give rise to a number
of interesting physical effects such as the doubling images of distant
objects or even gravitational lensing, the emission of gravitational waves
and the generation of high-energy cosmic rays (see, for instance, \cite {Damo00,Batta,BereAndHnatyk}).

The geometry of the spacetime associated with an idealized cosmic string,  i.e., infinitely long and straight,  is locally flat but topologically conical. It presents a planar angle deficit given by $\Delta\phi=8\pi G\mu_0$ \footnote{$G$ is the Newton's gravitational constant and $\mu_0$
is the linear mass density of the string.} on the two-surface orthogonal to the string. This object was first introduced in the literature as being created by a Dirac-delta type distribution of energy and axial stress. It can also be described by classical field theory where the energy-momentum tensor associated with the Maxwell-Higgs  system, investigated by Nielsen and Olesen in \cite{Nielsen197345}, couples to the Einstein's equations. This coupled system was first investigated in \cite{PhysRevD.32.1323} and \cite{Linet1987240}. In these papers, the authors have studied a planar angle deficit, $\Delta\phi $, arising on the two-surface perpendicular to a string, as well as a magnetic flux running through the string core.

It is well known that the vacuum in quantum field theory depends crucially on the presence of a background gauge field, and also on the topology and geometry of the spacetime where the theory is being defined. As a consequence, many physical observables may provide relevant informations about the structure of the physical system under investigation.  An important characteristic associated with charged fields is the vacuum expectation value (VEV) of the current density. Although the corresponding operator is local, because of the global nature of the quantum vacuum, this expectation value carries relevant informations about both the geometry and topology of the background spacetime. Moreover, this VEV acts as the source in the semiclassical Maxwell equations and therefore plays an important role in modeling a self-consistent dynamics involving the electromagnetic field. In the present paper we investigate the vacuum bosonic current induced by a magnetic flux running along an idealized cosmic string in an expanding Universe, represented by dS spacetime.
As we shall see, besides the natural dependence on the distance to the string, there appears an additional explicit time dependence on the current, reflecting the expansive character of the dS background. In addition we also admit that the coordinate along the string is compactified to a circle. As one knows, the compact spatial dimensions are an inherent feature of most high-energy theories of fundamental physics, including supergravity and superstring theories. In this way, we are investigating induced bosonic current in a dS space as general as possible. Obviously, when the size of the compact dimension tends to infinity we recover the results without compactification.

Due to its lack of global flatness, the cosmic string spacetime modifies the vacuum fluctuations associated with quantum fields. In this sense it has been shown non-vanishing results for the renormalized VEV  of physical observables, like the energy-momentum tensor, considering scalar and fermionic quantum fields in  \cite{PhysRevD.35.536,escidoc:153364, GL, DS, PhysRevD.46.1616} and \cite{PhysRevD.35.3779, LB, Moreira1995365, BK}, respectively. Moreover, if we take into consideration the presence of a magnetic flux running along the string's core, additional contributions to the above VEVs associated with charged fields take place. (See references in \cite{PhysRevD.36.3742, guim1994, SBM, SBM2, SBM3, Spinelly200477, SBM4}) Furthermore, the magnetic flux induces vacuum current densities, $\langle j^\mu\rangle$. This phenomenon has been investigated for massless and massive scalar fields in \cite{LS} and \cite{SNDV}, respectively. In these papers, 
the authors have shown that induced vacuum current densities along the azimuthal direction arise if the ratio of the magnetic flux by the quantum one has a nonzero fractional part. The calculation of induced bosonic and fermionic currents in higher-dimensional cosmic string spacetime in the presence of a magnetic flux has been developed in \cite{Eduardo,ERBM}. The induced fermionic current by a magnetic flux in $(2+1)$-dimensional conical spacetime and in the presence of a circular boundary has
also been analyzed in \cite{PhysRevD.82.085033}. In all the above-mentioned calculations the cosmic string was considered as an ideal linear object, i.e., without inner structure. The analysis of the scalar and fermionic vacuum current densities induced by a magnetic flux in a cosmic string considering a nonvanishing
core has been developed in \cite{Mello15} and  \cite{mikael}, respectively. Also, the calculation of the VEV of fermionic energy-momentum tensor appeared in \cite{mikael2}.

In general, the analysis of quantum effects associated with matter fields in a cosmic string spacetime, assume this defect in a flat background geometry. For a
cosmic string in a curved background, quantum effects associated with a
scalar field have been discussed in \cite{Davi88} for specific values of the
planar angle deficit. The vacuum polarization in Schwarzschild spacetime
pierced  by an infinite straight cosmic string was investigated in \cite{Otte10,Otte11}.
More recent publications studied the vacuum polarization effects induced by a cosmic string in a dS spacetime for scalar \cite{Beze09}, and fermionic 
 fields \cite{Beze10}. Similar analysis induced by a cosmic string in anti-dS spacetime, have been developed in \cite{Beze12} and \cite{Beze13} for massive scalar and fermionic fields, respectively. The calculation of vacuum fermionic current induced by a magnetic flux and
 the VEV of the energy-momentum tensor in dS spacetime in the presence of a compactified cosmic string, has been developed in \cite{mohammadi}
 and \cite{Braganca2019a}. Therefore, in the current paper we continue the same line of analysis by calculating the induced scalar current in the corresponding geometry. We also compare the results with the ones for a massive fermionic field obtained in \cite{mohammadi} in the same geometry.

This paper is organized as follows. In section \ref{Wightman} we describe the background geometry and construct the positive frequency Wightman function for a massive charged scalar quantum field in dS spacetime in the presence of a cosmic string presenting a magnetic flux along its axis. Moreover, we consider that the $z$-axis along the string is compactified to a circle and carries an extra magnetic flux. By using the Wightman function, in section \ref{current}, we evaluate the renormalized vacuum current
density induced by the magnetic fluxes and the compactification. The only nonzero components of the current densities correspond to the azimuthal and axial ones. These quantities are investigated in sections \ref{azimuthal} and \ref{axial}. The most relevant conclusions of the paper are summarized in section \ref{conclusion}.
We have also dedicated an Appendix to provide some important expressions used for the development of our study for the induced current densities.
Throughout the paper we use natural units $G=\hbar =c=1$.
%
\section{Wightman function}
\label{Wightman}
Let us consider a charged massive scalar field in a $(3+1)$ dS spacetime with an ideal cosmic string. We assume that the direction along the string is compactified to a circle with the length $L$.
The geometry associated with the corresponding background spacetime is given by the following line element
\begin{equation}
ds^{2}=dt^{2}-e^{2t/\alpha }(dr^{2}+r^{2}d\phi+dz^2)\ .  \label{ds1}
\end{equation}
The coordinates take values in the intervals $-\infty< t < +\infty$, $r\geq 0$, $0\leq\phi\leq \phi_0=2\pi/q$ and $0\leqslant z\leqslant L$.  The parameter $q\geq 1$ encodes the deficit angle due to the conical structure of the cosmic string, knowing that the minimum value $q=1$ shows its absence.  The parameter $\alpha$ is related to the cosmological constant and Ricci scalar through the expression,
$R=4\Lambda =12/\alpha ^{2}$.
For further analysis, in addition to the synchronous time coordinate $t$, it is more convenient to introduce the conformal
time $\tau $\ according to \cite{Beze09}
\begin{equation}
\tau =-\alpha e^{-t/\alpha }\ ,\quad  -\infty <\ \tau \ <\ 0\ .
\end{equation}%
leading to the following line element 
\begin{equation}
ds^{2}=(\alpha /\tau)^{2}(d\tau^{2}-dr^{2}-r^{2}d\phi^{2}-dz^2)\ .  \label{ds2}
\end{equation}
which shows a conformally flat spacetime.

In this paper we are interested in calculating the induced vacuum current density, $\langle j^{\mu}\rangle$, associated with
a charged scalar quantum field, $\varphi (x)$, in the presence of
magnetic flux running along the core of the string.
The equation which governs the quantum dynamics of a charged
bosonic field with mass $m$, interacting with an 
electromagnetic potential vector, $A_\mu$, in a curved background reads
\begin{equation}
\left({\cal D}^2+m^{2}+\xi R\right) \varphi (x)=0 \ , \quad \quad \textrm{with} \quad \quad {\cal D}^2=\frac{1}{\sqrt{|g|}}\, D_\mu\left(\sqrt{|g|}\,g^{\mu\nu}D_\nu\right) \ ,
\label{eq02}
\end{equation}
being $D_{\mu}=\partial_{\mu}+ieA_{\mu}$ and $g={\rm det}(g_{\mu\nu})$. In \eqref{eq02} we also considered
the presence of a non-minimal coupling, $\xi$, between the scalar field and the geometry represented by the
Ricci scalar $R$. Here, we are mainly interested in two specific values of the curvature coupling,  $\xi=0$ and $\xi=1/6$, which correspond
to the minimal and conformal coupling, respectively. It is known that for a bosonic field in a conformally flat spacetime only in the case of a massless scalar field with a conformal coupling $\xi=\frac{(D-1)}{4D}$ where D is the spatial dimension, the physical quantities are related with the corresponding flat one multiplied by some power of the conformal factor. 

The compactification along the $z$-axis is achieved by imposing the
quasiperiodicity condition on the matter field,
\begin{equation}
\varphi (t,r,\phi,z+L)=e^{2\pi i\beta}\varphi(t,r,\phi,z) \ ,  
\label{eq03}
\end{equation}
with a constant phase $\beta $, $0\leqslant \beta \leqslant 1$, where $\beta =0$ and $\beta =1/2$ correspond to the untwisted and twisted
fields, respectively, along the $z$-direction. For the rotation around the $z$-axis we adopt the periodic boundary condition
\begin{equation}
\varphi (t,r,\phi +\phi _{0},z)=\varphi(t,r,\phi,z) \ .  \label{eq03a}
\end{equation}

In addition, we consider the existence of a constant vector potential $A_{\mu}=(0,0,A_{\phi},A_{z})$ with $A_{\phi}=-q\Phi_\phi/(2\pi)$ and $A_{z}=-\Phi_z/L$, being $\Phi_\phi$ and $\Phi_z$ the corresponding magnetic fluxes. In quantum field  theory the  condition
\eqref{eq03} changes the spectrum of the vacuum fluctuations compared with the case of uncompactified dimension ($L \to \infty$) and, as a consequence, the induced vacuum current densities are dependent on the compactification parameter $L$, in general. 

The positive energy solution of \eqref{eq02} obeying the boundary conditions \eqref{eq03} and \eqref{eq03a} can be found in a similar way as it was calculated in \cite{Beze09} where the geometry of a straight cosmic string in dS spacetime, although without the magnetic flux and compactification, was considered. In the geometry we consider here, the normalized complete set of solutions generalizes to
\begin{eqnarray}
\label{sol1}
\varphi_\sigma(x)=\left(\frac{qp}{8 L\alpha^2}\right)^{1/2} e^{-i\pi \nu/2}\tau^{3/2}H_\nu^{(1)}(\lambda\tau)J_{q|n+a|}(pr) e^{inq\phi+ik_zz} \  , 
\end{eqnarray}
where $J_{\mu }(x)$ and $H_{\mu}^{(1)}(x)$ are the Bessel and Hankel functions \cite{Abra}, respectively, and 
\begin{eqnarray}
\lambda=\sqrt{p^2+{\tilde{k}}_z^2} \ , \ {\rm with } \ {\tilde{k}}_z=k_z+eA_z \ , \ p\geq 0 \ .
\label{eq09}
\end{eqnarray}
The parameters in the order of Hankel and Bessel functions are
\begin{eqnarray}
\nu=\frac12\sqrt{9-48\xi-4m^2\alpha^2}  \, , \quad \quad  a=\frac{eA_\phi}{q}=-\frac{\Phi_\phi}{\Phi_0} 
\label{parama}
\end{eqnarray}
with  $\Phi_0=2\pi/e$ being the quantum flux.

The quasiperiodicity condition \eqref{eq03} results in a discretization 
of the quantum number $k_z$ as shown below
\begin{equation}
k_z=k_l=\frac{2\pi}{L}(l+\beta)  \quad {\rm with} \quad  l=0\ ,\pm 1,\pm 2,... \ ,
\label{eq11}
\end{equation}
which gives
\begin{eqnarray}
{\tilde{k}}_z={\tilde{k}}_l=\frac{2\pi}{L}(l+\tilde{\beta})  \quad {\rm with}
\quad \tilde{\beta}=\beta+\frac{eA_zL}{2\pi}=\beta-\frac{\Phi_z}{\Phi_0} \ .
\label{eq13}
\end{eqnarray}
As a result, the positive-energy solution \eqref{sol1} is characterized by the set of quantum numbers, $\sigma=\{p,n,l\}$. 

In order to find the induced bosonic current densities we need first to calculate the  Wightman function which describes the properties of the vacuum state. The positive 
frequency Wightman function is given by $W(x,x')=\left\langle 0|{\hat{\varphi}}(x){\hat\varphi}^{*}(x')|0 \right\rangle$, 
where $|0 \rangle$ stands for the vacuum state and $\hat{\varphi}(x)$ the field operator. For the evaluation of the Wightman 
function, we use the mode sum formula
\begin{equation}
W(x,x')= \sum_{n=-\infty}^{+\infty}  \ \int_0^\infty
\, dp \ \sum_{l=-\infty }^{+\infty} \varphi_{\sigma}(x)\varphi_{\sigma}^{*}(x') \ .
\label{eq17}
\end{equation}
The set $\{\varphi_{\sigma}(x), \ \varphi_{\sigma}^{*}(x')\}$ represents a complete set of normalized mode functions satisfying the quasiperiodicity condition \eqref{eq03}.

Substituting \eqref{sol1} into the sum \eqref{eq17}, and after several intermediate steps we obtain
\begin{align}
W(x,x')&=\frac{2q(\tau\tau')^{3/2}e^{-ieA_z(z-z')}} {(2\pi\alpha)^2 L} \int_0^\infty dp \, p \sum_{l=-\infty}^{\infty}e^{\frac{2\pi}Li(l+\tilde{\beta)}(z-z')}\nonumber\\
&\times\sum_{n=-\infty}^{\infty}e^{iqn(\phi-\phi')}J_{q|n+a|}(pr)J_{q|n+a|}(pr')
K_\nu(i\lambda\tau')K_\nu(-i\lambda\tau) \  .
\label{eq18}
\end{align}
To obtain the above expression we have used the relation between the Hankel function, $H_\nu^{(1)}(z)$, and the modified Bessel function with imaginary argument, $K_\mu(-iz)$ \cite{Abra}. In order to simplify the summation over $l$, we have introduced an exponential function $e^{-ieA_z(z-z')}$ to replace the exponent factor $\beta$ by $\tilde{\beta}$, using relation \eqref{eq13}, in the summation over the quantum number $l$. 
With the objective to develop the summation over the quantum number $l$ we apply
the Abel-Plana summation formula \cite{PhysRevD.82.065011} in the form below
\begin{align}
\sum_{l=-\infty }^{\infty }g(l+\tilde{\beta} )f(|l+\tilde{\beta}|)&=\int_{0}^{\infty }du\,
\left[ g(u)+g(-u)\right] f(u)  \notag \\
& +i\int_{0}^{\infty }du\left[ f(iu)-f(-iu)\right] \sum_{\chi =\pm
	1}\frac{g(i\chi u)}{e^{2\pi (u+i\chi \tilde{\beta} )}-1} \ .
\label{sumform}
\end{align}

For the case under investigation, we consider $u\equiv \frac{L}{2\pi}k_z$ as well as
\begin{eqnarray}
\label{gf}
g(l+\tilde{\beta} )=e^{\frac{2\pi}Li(l+\tilde{\beta)}(z-z')}  \quad \quad \textrm{and} \quad \quad
f(|l+\tilde{\beta} |)=K_\nu(i\lambda\tau')K_\nu(-i\lambda\tau) \ ,
\end{eqnarray}
where
\begin{equation}
\lambda=\lambda_l=\sqrt{\left({2\pi}(l+\tilde{\beta})/L)\right)^2+p^2} \ .
\end{equation} 
At this point, we can express the Wightman function as the sum of two terms as shown below
\begin{eqnarray}
\label{w-tot}
W(x,x')=W_{s}(x,x')+W_c(x,x') \  , 
\end{eqnarray}
where the first contribution, $W_{s}(x,x')$, represents the Wightman function in dS spacetime in the presence of an uncompactified cosmic string, and the second, $W_{c}(x,x')$, is induced by the compactification. Let us call the former and the latter, the string and the compactified Wightman functions, respectively.

Substituting \eqref{gf} into \eqref{sumform}, and also substituting the result into \eqref{eq18} we get
\begin{eqnarray}
\label{wightcs}
W_{s}(x,x')&=&\frac{2q(\tau\tau')^{3/2}e^{-ieA_z(z-z')}} {(2\pi)^3\alpha^2}\int dk_z \ e^{ik_z \Delta z}\sum_{n=-\infty}^{\infty}e^{iqn(\phi-\phi')}\int_0^\infty dp \, p \nonumber\\
&&\times \, J_{q|n+a|}(pr)J_{q|n+a|}(pr')
K_\nu\left(i\tau'\sqrt{p^2+k_z^2}\right)K_\nu\left(-i\tau\sqrt{p^2+k_z^2}\right) \  ,
\end{eqnarray}	
the string Wightman function, originating from the first term on the right hand side of \eqref{sumform}.

The string Wightman function \eqref{wightcs} can be expressed in terms of a more workable integral representation in a similar way as developed in Appendix A of \cite{Beze09}. The final expression is,
\begin{align}
\label{wightcs1}
W_{s}(x,x')&=\frac{q(\tau\tau')^{3/2}e^{-ieA_z\Delta z}} {2\pi^{5/2}\alpha^2}\sum_{n=-\infty}^\infty e^{iqn\Delta\phi}\int_0^\infty dx \,
x^{1/2}e^{-{\cal{V}} x} \, I_{q|n+a|}(2xrr')K_\nu(2\tau'\tau x) \  ,
\end{align}
with 
\begin{eqnarray}
{\cal{V}}=(z-z')^2+r^2+r'^2-(\tau^2+\tau'^2) \  .
\end{eqnarray}

As to the compactified Wightman function, it is given by the second term on the right hand side of \eqref{sumform}. In Appendix \ref{compacified} we show that $W_c(x,x')$ can be expressed in the following form
\begin{eqnarray}
\label{wight2}
W_c(x,x')&=&\frac{2q(\tau\tau')^{3/2}e^{-ieA_z\Delta z}} {(2\pi\alpha)^2}\sum_{n=-\infty}^{\infty}e^{iqn(\phi-\phi')}\int_0^\infty dp \, p \, 
J_{q|n+a|}(pr)J_{q|n+a|}(pr')\nonumber\\
&&\times \, \int_0^\infty\frac{d\gamma\, \gamma}{\sqrt{\gamma^2+p^2}}\left[K_\nu(\tau'\gamma)I_{-\nu}(\tau\gamma)+
I_\nu(\tau'\gamma)K_\nu(\tau\gamma)\right]\nonumber\\
&&\times \, \sum_{l=1}^\infty e^{-lL\sqrt{\gamma^2+p^2}}\cos\left(2\pi l\tilde{\beta}-i\Delta z\sqrt{\gamma^2+p^2}\right) \  .
\end{eqnarray}
With the Wightman functions \eqref{wightcs1} and \eqref{wight2} in hands, we are in position to
evaluate the induced current densities. 
%
\section{Bosonic current}
\label{current}
%
The bosonic current density operator is given by
\begin{eqnarray}
\hat{j}_{\mu }(x)&=&ie\left[{\hat\varphi} ^{*}(x)D_{\mu }{\hat\varphi} (x)-
(D_{\mu }{\hat\varphi})^{*}{\hat\varphi}(x)\right] \nonumber\\
&=&ie\left[{\hat\varphi}^{*}(x)\partial_{\mu }{\hat\varphi} (x)-\varphi(x)
(\partial_{\mu }{\hat\varphi}(x))^{*}\right]-2e^2A_\mu(x)|{\hat\varphi}(x)|^2 \   .
\label{eq20}
\end{eqnarray}
Its vacuum expectation value (VEV) can be evaluated in terms of the positive frequency Wightman function as follows
\begin{equation}
\left\langle j_{\mu}(x) \right\rangle=ie\lim_{x'\rightarrow x}
\left\{(\partial_{\mu}-\partial_{\mu'})W(x,x')+2ieA_\mu W(x,x')\right\} \ .
\label{eq21}
\end{equation}

Due to the decomposition \eqref{w-tot}, the current density can be written as,
\begin{eqnarray}
\left\langle j_{\mu}(x) \right\rangle=\left\langle j_{\mu}(x) \right\rangle_{s}+\left\langle j_{\mu}(x) \right\rangle_c \  ,
\label{decomp}
\end{eqnarray}
where the first term corresponds to the contribution of an uncompactified cosmic string in dS spacetime, while
the second one is the contribution induced by the compactification. 
As we will see shortly, the non-vanishing components of the current density are periodic functions of the magnetic fluxes $\Phi_\phi$ and $\Phi_z$ with the period equal to the quantum flux. This can be observed easily writing  the parameter $a$ in \eqref{parama} in the form $a = n_{0}+a_{0}  \ {\rm with} \ |a_{0}|<\frac{1}{2}$,
where $n_{0}$ is an integer number. In this case the VEV of the current density depends on $a_{0}$ only.

In the problem under consideration, the only nonzero components of the current density are the azimuthal and axial ones.  
%
\subsection{Azimuthal current}
\label{azimuthal}
%
The VEV of the azimuthal current density is given by
\begin{equation}
\left\langle j_{\phi}(x) \right\rangle = ie \lim_{x '\rightarrow x}
\left\{(\partial_{\phi}-\partial_{\phi '})W(x,x')+2ieA_{\phi}W(x,x')\right\} \ .
\label{jphi}
\end{equation}
The above expression can be decomposed as in \eqref{decomp}.
Let us start with the string contribution.
Substituting \eqref{wightcs1} into \eqref{jphi} as well as taking the derivative and coincidence limit, one finds
\begin{align}
\left\langle j_{\phi}(x) \right\rangle_{s}&=-\frac{q^2e\tau^3}{\pi^{5/2}\alpha^2}
\int_0^\infty dx \, x^{1/2}e^{-2(r^2-\tau^2)x}K_\nu(2\tau^2x)\mathcal{J}(q,a_0,2xr^2)\ ,
\label{jphics1}
\end{align}
where
\begin{equation}
\mathcal{J}(q,a_0,2xr^2)=\sum_{n=-\infty}^\infty(n+a_0)I_{q|n+a_0|}(2xr^2)\ .
\end{equation}
The summation above has been developed in \cite{Eduardo}. The result is 
\begin{align}
\mathcal{J}(q,a_0,w)&=\frac{2w}{q^2}\sideset{}{'}\sum_{k=1}^{[q/2]}\sin(2k\pi/q)\sin(2k\pi a_0)
e^{w\cos(2k\pi/q)}+\frac{w}{q\pi}\int_0^\infty dy \, \sinh y\frac{e^{-w\cosh y}g(q,a_0,y)}{\cosh(qy)-\cos(q\pi)}\ ,\nonumber\\
\label{Summation2}
\end{align}
with
\begin{equation}
g(q,a_0,y)=\sin(q\pi a_0)\sinh[(1-|a_0|)qy]-\sinh(qya_0)\sin[(1-|a_0|)q\pi].
\end{equation}
In the expression \eqref{Summation2}, $[q/2]$ means the integer part of this ratio and the prime in the summation
symbol means that in the case of $q/2$ is an integer, the term should be taken with the coefficient $1/2$.

Taking into account \eqref{Summation2},  the Eq. \eqref{jphics1} becomes
\begin{align}
\left\langle j_{\phi}(x) \right\rangle_{s}&=-\frac{e}{2^{1/2}\pi^{5/2}}\left(\frac{r}{\alpha\tau}\right)^2
\int_0^\infty dz \,z^{3/2}K_\nu(z)\left[\sideset{}{'}\sum_{k=1}^{[q/2]}\sin(2k\pi/q)\sin(2k\pi a_0)\,  e^{-\left(\frac{2r^2s_k^2}{\tau^2}-1\right)z}
\right.\nonumber\\
&\left. +\frac{q}{\pi}\int_0^\infty dy \, \frac{\sinh (2y)g(q,a_0,2y)}{\cosh(2qy)-\cos(q\pi)}
e^{-\left(\frac{2r^2c_y^2}{\tau^2}-1\right)z}\right],
\label{jphics2}
\end{align}
where we have introduced $z=2\tau^2x$ and defined
\begin{equation}
s_k\equiv\sin(k\pi/q) \ \ \ {\rm and} \ \ \ c_y\equiv\cosh(y).
\end{equation}
The integral over $z$ can be evaluated with the help of the formula \cite{Beze09}
\begin{eqnarray}
\int_0^\infty dx \,x^{(\mu-3)/2}e^{-cx}K_\nu(\beta x)&=&\frac{2^{1-\mu}\sqrt{\pi}}{\Gamma(\mu/2)(\beta/2)^{(\mu-1)/2}}
\Gamma\left(\frac{\mu-1}{2}-\nu\right)\Gamma\left(\frac{\mu-1}{2}+\nu\right)\nonumber\\
&&\times \,F\left(\frac{\mu-1}{2}+\nu,\frac{\mu-1}{2}-\nu,\mu/2;\frac{1}{2}-\frac{c}{2\beta}\right),
\end{eqnarray}
with $F(a,b,c;x)$ being the hypergeometric function. After the evaluation of the integral
over $z$ the final expression for the azimuthal current induced
by the cosmic string in dS background is given by
\begin{align}
\left\langle j^{\phi}(x) \right\rangle_{s}&=\frac{e}{(4\pi)^{2}\alpha^4}
\left[\sideset{}{'}\sum_{k=1}^{[q/2]}\sin(2k\pi/q)\sin(2k\pi a_0) \, \mathcal{B}_{0}\left(r/\tau,s_k\right)\right.\nonumber\\
&\left.+\frac{q}{\pi}\int_0^\infty dy \, \frac{\sinh (2y)g(q,a_0,2y)}{\cosh(2qy)-\cos(q\pi)} \, \mathcal{B}_{0}\left(r/\tau,c_y
\right)\right],
\label{jphicsFinal}
\end{align}
which is an odd function of the magnetic flux $a_0$ and depends on the ratio $r/\tau$. This ratio is
the proper distance from the string in units of the dS curvature ratio $\alpha$. In addition, we have used the definition
\begin{equation}
\mathcal{B}_{l}(u,v)\equiv \Gamma\left(\frac{5}{2}+\nu\right)\Gamma\left(\frac{5}{2}-\nu\right)\, 
F\left(\frac{5}{2}+\nu,\frac{5}{2}-\nu;3;1-u^2v^2-\frac{l^2L^2}{4\tau^2}\right),
\label{Fdefinition}
\end{equation}
with $l=0$.
The above definition will also be useful in the rest of the calculations.
Notice that the parameter $\nu$, defined in \eqref{parama}, is related to the mass of the scalar field and can be a real or an imaginary quantity. 

 At this point we would like to analyze some asymptotic cases of the azimuthal current density.  Let us first consider the situation where, $r/\tau \gg 1$. For a fixed conformal time this corresponds to the region far from the string.  In this region the  the main contribution to the integral over $z$ in \eqref{jphics2} comes from small values of the argument of the Macdonald function. By considering the corresponding expansion of the Macdonald function for small arguments and
	$\nu$ as imaginary, we find 
\begin{align}
\left\langle j^{\phi}(x) \right\rangle_{s}&\approx\frac{e}{8\pi^{5/2}\alpha^4}\left(\frac{\tau}{r}\right)^{5}
\left\{\sideset{}{'}\sum_{k=1}^{[q/2]}\frac{\sin(2k\pi/q)\sin(2k\pi a_0)}{s_k^{5}} \,
{\rm Re}\left[\left(\frac{2rs_k}{\tau}\right)^{2\nu}\Gamma(\nu)\Gamma(5/2-\nu)\right]\right.\nonumber\\
&\left. + \, \frac{q}{\pi}\int_0^\infty dy \, \frac{\sinh (2y)g(q,a_0,2y)c_y^{-5}}{\cosh(2qy)-\cos(q\pi)} 
\, {\rm Re}\left[\left(\frac{2rc_y}{\tau}\right)^{2\nu}\Gamma(\nu)\Gamma(5/2-\nu)\right]\right\}.
\label{jphicsLargeR}
\end{align}
From the above expression we note an oscillatory behavior at large distances from the string when $\nu$ is imaginary. As can be seen, for large distances the string contribution of the azimuthal current density tends to zero with the fifth power of $\tau/r$. If $\nu$ is real, we have the same equation, \eqref{jphicsLargeR}, with the coefficient $1/2$. However, knowing that for the case of a minimal coulpling $(\xi=0),$ $0\le m\alpha \le 3/2$ for real $\nu$, depending on the mass of the bosonic field, the decay behavior for large distances can be really different. As $\nu$ becomes closer to the maximum value (as mass decreases to zero), the decay in $\left\langle j^{\phi}(x) \right\rangle_{s}$ becomes slower. It tends to zero with the second power of $\tau/r$ for the massless field. In contrast, $\left\langle j^{\phi}(x) \right\rangle_{s}$ for the fermionic field in the same geometry decays with the fourth power of $\tau/r$ independent of the mass, and while decaying oscillates as it was shown in \cite{mohammadi}. However, in the massless limit the oscillating behavior disappears.
Now, in the case of a conformal coulpling, $\xi=1/6$ resulting in $0\le m\alpha \le 1/2$ for real $\nu$, the azimuthal current density
goes to zero with the fourth power of $\tau/r$ in the massless limit, similar to the fermionic case.

Moreover, by explicit derivation we can show that in the limit $r/\tau\ll1$, i.e., in the regions near the string, Eq. \eqref{jphicsFinal} presents a divergence on the string with the inverse fourth power of the proper distance matching exactly with the result for the fermionic field studied in \cite{mohammadi}.

\begin{figure}[h]
	\centering
	{\includegraphics[width=0.48\textwidth]{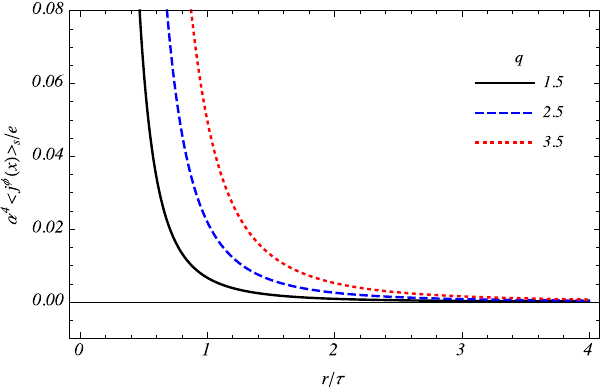}}
	\hfill
	{\includegraphics[width=0.49\textwidth]{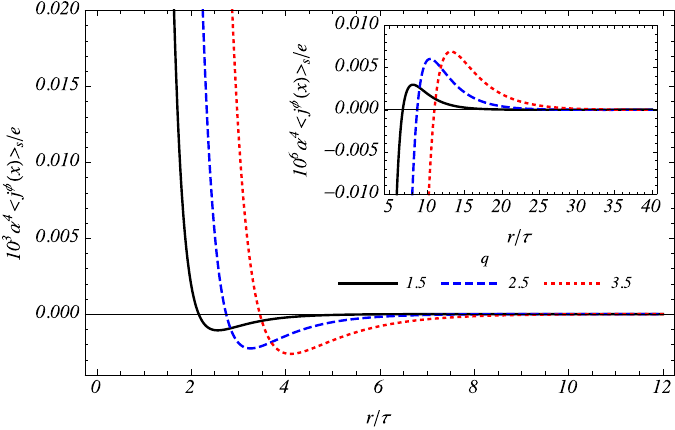}}
	\caption{String part of the azimuthal current density as a function of $r/\tau$ considering several values of the parameter $q$. In the left plot
	$m\alpha=1$ while in the right one $m\alpha=2$. In both plots $a_0=0.25$ and $\xi=0$.}
	\label{fig01}
\end{figure}

In Fig. \ref{fig01} we plot the string part of the azimuthal current density (in the absence of the compactification) as a function
of $r/\tau$ considering different values of $q$ and a minimal coupling ($\xi=0$). In the left plot, the parameter $\nu$ is real and in the right one is imaginary.
In the latter case we show the oscillatory behavior of the azimuthal current density at large distances from the string.

Now, for the evaluation of the azimuthal current density induced by the compactification, we need to start again with the
Eq. \eqref{wight2}. Substituting this equation into \eqref{jphi}, after taking the derivatives and
the coincidence limit, one obtains
\begin{eqnarray}
\left\langle j_{\phi}(x) \right\rangle_{c}&=&-\frac{4eq^2\tau^3}{(2\pi\alpha)^2}\sum_{l=1}^\infty
\cos(2\pi l\tilde{\beta})\sum_{n=-\infty}^\infty(n+a_0)\int_0^\infty dp \, p \, J^2_{q|n+a|}(pr)\nonumber\\
&&\times\int_0^\infty d\gamma \, \gamma \, \frac{e^{-lL\sqrt{\gamma^2+p^2}}}{\sqrt{\gamma^2+p^2}}
K_\nu(\tau\gamma)[I_\nu(\tau\gamma)+I_{-\nu}(\tau\gamma)].
\end{eqnarray}
After evaluating the integrals over $p$ and $\gamma$  \cite{gradshteyn2000table,PhysRevD085002}, the expression becomes
\begin{align}
\left\langle j_{\phi}(x) \right\rangle_{c}&=-\frac{2eq^2\tau^3}{\alpha^2\pi^{5/2}}\sum_{l=1}^\infty
\cos(2\pi l\tilde{\beta})\int_0^\infty dx \, x^{1/2}e^{-(l^2L^2+2r^2-2\tau^2)x} \, K_\nu(2\tau^2x)\mathcal{J}(q,a_0,2r^2x),
\end{align}
where $x=1/(4s^2)$. In the above equation, it is not difficult to see that in the limit $L\to \infty$ the contribution of the current originated from the compactification vanishes which is the expected result. By using the representation \eqref{Summation2} and after the integration over
$x$, one finds
\begin{eqnarray}
\left\langle j^{\phi}(x) \right\rangle_{c}&=&\frac{2e}{(4\pi)^2\alpha^4}\sum_{l=1}^\infty
\cos(2\pi l\tilde{\beta})\left[\sideset{}{'}\sum_{k=1}^{[q/2]}\sin(2k\pi/q)\sin(2k\pi a_0) \, \mathcal{B}_{l}\left(r/\tau,s_k\right)\right.\nonumber\\
&&\left.+\, \frac{q}{\pi}\int_0^\infty dy \, \frac{\sinh (2y)g(q,a_0,2y)}{\cosh(2qy)-\cos(q\pi)} \, \mathcal{B}_{l}\left(r/\tau,c_y
\right)\right],
\label{jphicompFinal}
\end{eqnarray}
where we have used the definition \eqref{Fdefinition}. The above expression is an even function of the magnetic flux enclosed by
the compactification and an odd function of the magnetic flux running along the string.

Some asymptotic cases can be studied for the current density induced by the compactification.
Considering small distances from the string, we can use
the following expression for the hypergeometric function \cite{Abra}
\begin{eqnarray}
F(a,b,c;z)&=&\frac{\Gamma(c)\Gamma(b-a)}{\Gamma(b)\Gamma(c-a)}
\frac{F\left(a,c-b,a-b+1;\frac{1}{1-z}\right)}{(1-z)^a}\nonumber\\
&&+\, \frac{\Gamma(c)\Gamma(a-b)}{\Gamma(a)\Gamma(c-b)}
\frac{F\left(b,c-a,b-a+1;\frac{1}{1-z}\right)}{(1-z)^b}.
\label{formulaF}
\end{eqnarray}
For the case where the parameter $\nu$ is imaginary, in the leading term we have
\begin{eqnarray}
\left\langle j^{\phi}(x) \right\rangle_{c}&\approx&\frac{eq\sin(q\pi a_0)}{(2\pi)^3\alpha^4}
\left(\frac{\tau}{r}\right)^{2(1-q|a_0|)}\Gamma(1-q|a_0|)\sum_{l=1}^\infty\cos(2\pi l\tilde{\beta})\nonumber\\
&&\times \, {\rm Re}\left[\frac{\Gamma(2\nu)\Gamma\left(\frac{3}{2}+q|a_0|-\nu\right)}
{\Gamma\left(\frac{1}{2}+\nu\right)}
\left(\frac{2\tau}{lL}\right)^{2(q|a_0|-1-\nu)+5}\right].
\label{jCompSmallR}
\end{eqnarray}
If $\nu$ is real, the leading contribution comes from the second term on the right-hand side of \eqref{formulaF} where the result is $1/2$ of the imaginary $\nu$ case.
In this limit, the VEV of the azimuthal current density induced by the compactification diverges on the string if $|a_0|\leq1/q$ and
is finite on the string if $|a_0|>1/q$, unlike the fermionic case where the threshold value is $|a_0|=(1-1/q)/2$ \cite{mohammadi}.
Now, in the limit where $L/\tau \gg 1$ and $r/\tau$ fixed, again we can use \eqref{formulaF} and for $\nu$ being imaginary one finds
\begin{eqnarray}
\left\langle j^{\phi}(x) \right\rangle_{c}&\approx&\frac{e}{2\pi^2\alpha^4}\left(\frac{\tau}{L}\right)^5
\sum_{l=1}^\infty\cos(2\pi l\tilde{\beta})\left[\sideset{}{'}\sum_{k=1}^{[q/2]}\sin(2k\pi/q)\sin(2k\pi a_0)\, 
\mathcal{G}_l(r/L,s_k)\right.\nonumber\\
&&\left.+\, \frac{q}{\pi}\int_0^\infty dy \, \frac{\sinh (2y)g(q,a_0,2y)}{\cosh(2qy)-\cos(q\pi)} \, \mathcal{G}_{l}\left(r/L,c_y
\right) \right],
\label{jphicLargL}
\end{eqnarray}
where we use the notation
\begin{equation}
\mathcal{G}_l(u,v)\equiv{\rm Re}\left[\frac{\Gamma(2\nu)\Gamma(5/2-\nu)}{\Gamma(1/2+\nu)}
\left(\frac{L}{\tau}\right)^{2\nu}\left(\frac{l^2}{4}+u^2v^2\right)^{\nu-5/2}\right].
\end{equation}
Clearly, the compactification contribution of the azimuthal current density tends to zero when $L\to \infty$.

\begin{figure}[h]
	\centering
	{\includegraphics[width=0.49\textwidth]{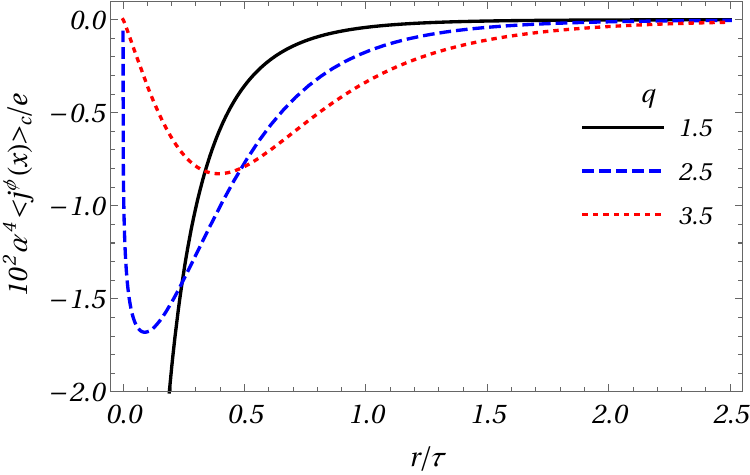}}
	\hfill
	{\includegraphics[width=0.49\textwidth]{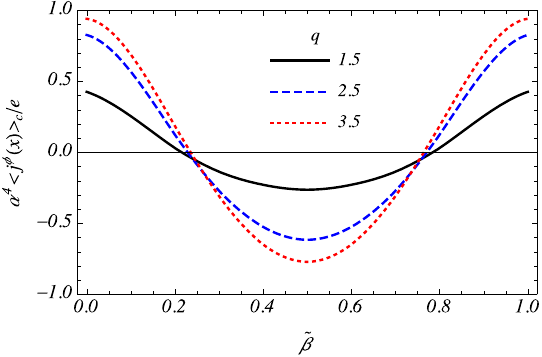}}
	\caption{Azimuthal current density induced by the compactification as a function of $r/\tau$ (left plot) and $\tilde{\beta}$
	(right plot) for different values of the parameter $q$. In the left
	plot we consider $\tilde{\beta}=0.25$, $a_0=0.45$ and in the right one $r/\tau=a_0=0.25$. In both plots
	we have $m\alpha=1.5$, $L/\tau=0.75$ and $\xi=0$.}
	\label{fig03}
\end{figure}

In Fig. \ref{fig03}, taking into consideration a minimal couping, we plot the VEV of the azimuthal current density induced by the compactification. In the left
plot we have the induced current as a function of $r/\tau$ for three different values of $q$. Note that the the left plot shows both cases where the azimuthal current density diverges or not on the string depending on the value of $q$. Moreover, when $r/\tau\to \infty$ it goes to zero. In the right plot of Fig. \ref{fig03}
the azimuthal current density induced by the compactification is shown as a function of $\tilde{\beta}$ for the same values of the parameter $q$. It exhibits the periodic behavior in $\tilde{\beta}$, as expected.

We note that taking into account the Eqs. \eqref{jphicsFinal} and \eqref{jphicompFinal} is
possible to write the total azimuthal current density as
\begin{eqnarray}
\left\langle j^{\phi}(x) \right\rangle&=&\frac{2e}{(4\pi)^2\alpha^4}\sideset{}{'}\sum_{l=0}^\infty
\cos(2\pi l\tilde{\beta})\left[\sideset{}{'}\sum_{k=1}^{[q/2]}\sin(2k\pi/q)\sin(2k\pi a_0) \, \mathcal{B}_{l}\left(r/\tau,s_k\right)\right.\nonumber\\
&&\left.+\, \frac{q}{\pi}\int_0^\infty dy \, \frac{\sinh (2y)g(q,a_0,2y)}{\cosh(2qy)-\cos(q\pi)} \, \mathcal{B}_{l}\left(r/\tau,c_y
\right)\right].
\label{jphitotal}
\end{eqnarray}
The term $l=0$ has to be  taken with the coefficient $1/2$ and give us the azimuthal current density induced by the uncompactified string and the curvature of the dS spacetime. For a massless scalar field and considering the conformal coupling, $\xi=1/6$, we have $\nu=1/2$. For this specific case, by using the properties of the hypergeometric function \cite{Abra}, the total azimuthal current matches exactly to the one for the flat spacetime in the presence of a compactified cosmic string \cite{Eduardo},  for the $(1+3)-$dimension,  times the conformal factor $(\tau/\alpha)^4$ which is the expected result.

\subsection{Axial current}
\label{axial}
Now, we focus on the axial current density, which its VEV is given by
\begin{equation}
\left\langle j_{z}(x) \right\rangle = ie \lim_{x '\rightarrow x}
\left\{(\partial_{z}-\partial_{z '})W(x,x')+2ieA_{z}W(x,x')\right\} \ .
\label{jz}
\end{equation}
Replacing the Wightman function \eqref{wightcs1} in the above equation, after taking the derivatives and the coincidence limit
we note that the string contribution vanishes. 

The contribution of the axial current density induced by the compactification after several intermediate steps, similar to the ones done in the last sections,
is given in the following form 
\begin{eqnarray}
\left\langle j_{z}(x) \right\rangle_{c}&=&-\frac{eq\tau^3}{(\pi\alpha)^2}
\sum_{l=1}\sin(2\pi l\tilde{\beta})\sum_{n=-\infty}^\infty
\int_0^\infty dp \, p J^2_{q|n+a|}(pr)\int_0^\infty d\gamma \, \gamma K_\nu(\nu\gamma)\nonumber\\
&\times&[I_\nu(\tau\gamma)+I_{-\nu}(\tau\gamma)]e^{-lL\sqrt{\gamma^2+p^2}}.
\label{jzc1}
\end{eqnarray}
The integral representation
\begin{equation}
e^{-lL\sqrt{\gamma^2+p^2}}=\frac{lL}{\sqrt{\pi}}\int_0^\infty ds \, s^{-2}\, e^{-(\gamma^2+p^2)s^2-l^2L^2/(4s^2)}.
\end{equation}
allows us to perform the integration over $p$ and $\gamma$ in the
expression \eqref{jzc1} where the result is given by
\begin{align}
\left\langle j_{z}(x) \right\rangle_{c}&=-\frac{4eqL\tau^3}{\alpha^2\pi^{5/2}}
\sum_{l=1}^\infty l\sin(2\pi l\tilde{\beta})\int_0^\infty dx \, x^{3/2}e^{-(l^2L^2+2r^2-2\tau^2)x}K_{\nu}(2\tau^2x)
\mathcal{I}(q,a_0,2r^2x),
\end{align}
with $x \equiv 1/(4s^2)$ and
\begin{eqnarray}
{\cal{I}}(q,a_0,w)=
\sum_{n=-\infty}^\infty I_{q|n+a_0|}(w) \ .
\label{sum1}
\end{eqnarray}
In \cite{Eduardo}, it has been shown that \eqref{sum1} can be written as
\begin{align}
\label{sumform1}
{\cal{I}}(q,a_{0},w)&=\frac{e^w}{q}-\frac{1}{\pi}
\int_{0}^{\infty}dy\frac{e^{-w\cosh y}f(q,a_0,y)}{\cosh(qy)-\cos(\pi q)}+\frac{2}{q}\sideset{}{'}\sum_{k=1}^{[q/2]}\cos(2k\pi a_0)e^{w\cos(2k\pi/q)} \ ,
\end{align}
where $[q/2]$ represents the integer part of $q/2$, and the prime on the sign of
the summation means that in the case $q=2p$ the term $k=q/2$ should be
taken with the coefficient $1/2$.
Obviously, if $q<2$ the summation term must be omitted.
Moreover, the function $f(q,a_0,y)$ is given by
\begin{eqnarray}
f(q,a_0,y)=\sin[(1-|a_0|)\pi q]\cosh(|a_0| qy)+
\sin(|a_0|\pi q)\cosh[(1-|a_0|)qy]    \  .
\label{func}
\end{eqnarray}
After
the integration over $x$ \cite{gradshteyn2000table}, the final expression for the axial current density is given by 
\begin{eqnarray}
\left\langle j^{z}(x)\right\rangle_{c}&=&\frac{eL}{2(2\pi)^2\alpha^4}\sum_{l=1}^\infty l\sin(2\pi l\tilde{\beta})
\left[\sideset{}{'}\sum_{k=0}^{[q/2]}\cos(2k\pi a_0) \, \mathcal{B}_{l}\left(r/\tau,s_k\right)\right.\nonumber\\
&&\left. \, -\frac{q}{\pi}\int_0^\infty dy \, \frac{f(q,a_0,2y)}{\cosh(2qy)-\cos(q\pi)} \, \mathcal{B}_{l}\left(r/\tau,c_y
\right)\right].
\label{jzc2}
\end{eqnarray}

For the case of a massless scalar field and considering a conformal coupling, we have $\nu=1/2$. By similar considerations done for the azimuthal current, we find that the axial current density matches to the result for the Minkowski spacetime times the conformal factor $(\tau/\alpha)^4$ (the sign of the axial current obtained in \cite{Eduardo} should be corrected to the opposite one).

Considering the term $k=0$ of the Eq. \eqref{jzc2} we have
\begin{eqnarray}
\left\langle j^{z}(x) \right\rangle_{c}^{(0)}&=&\frac{eL}{\pi^2(2\alpha)^4}\sum_{l=1}^\infty l\sin(2\pi l\tilde{\beta}) \, 
\mathcal{B}_{l}\left(r/\tau,s_0\right).
\label{jzk0}
\end{eqnarray}
This expression is the VEV of the axial current density with no dependence on $q$ and $a_0$. It is a
purely topological term being dependent only on the compactification of the $z-$axis and obviously goes to zero in the limit $L\to \infty$. The remaining in \eqref{jzc2}, magnetic flux and planar angle deficit contribution,                                                                                                                                                                                                                                                                                                                                                                                                                                                                                                                                                                                                                                                                                                                                                                                                                                                                                                                                                                                                                                                                                                                                  in the axial current density is given by
\begin{eqnarray}
\left\langle j^{z}(x) \right\rangle_{c}^{(q,a_0)}&=&\frac{eL}{2(2\pi)^2\alpha^4}\sum_{l=1}^\infty l\sin(2\pi l\tilde{\beta})
\left[\sideset{}{'}\sum_{k=1}^{[q/2]}\cos(2k\pi a_0) \, \mathcal{B}_{l}\left(r/\tau,s_k\right)\right.\nonumber\\
&&\left.-\frac{q}{\pi}\int_0^\infty dy \, \frac{f(q,a_0,2y)}{\cosh(2qy)-\cos(q\pi)} \, \mathcal{B}_{l}\left(r/\tau,c_y
\right)\right].
\label{jzc3}
\end{eqnarray}
This expression  is finite on the string. Besides that, it is an odd function of the magnetic flux enclosed by the compactification along the
string axis and an even function of the magnetic flux along the string's core.

\begin{figure}[h]
	\centering
	{\includegraphics[width=0.49\textwidth]{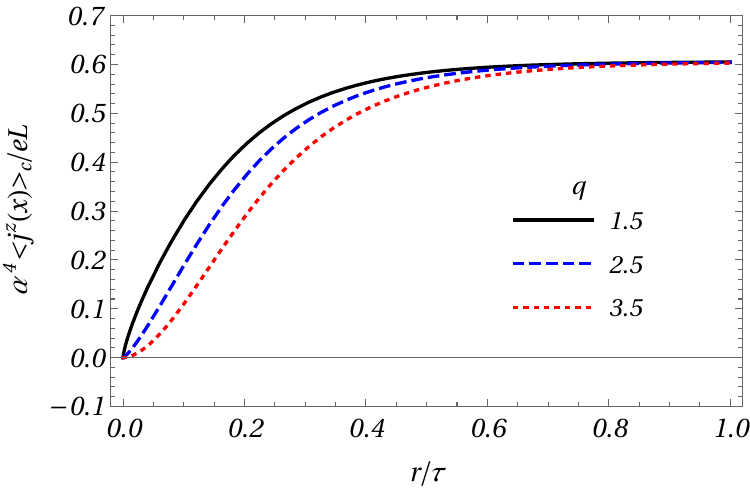}}
	\hfill
	{\includegraphics[width=0.49\textwidth]{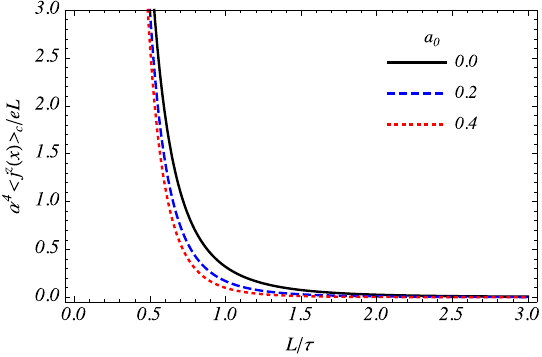}}
	\caption{Axial induced current density as a function of $r/\tau$ (left plot) for three values of $q$
	and as a function of $L/\tau$ (right plot) for three different values of $a_0$. In the left graph $L/\tau=0.75$,  
	while in the right one $r/\tau=0.25$ and $q=1.5$.  In both plots $\tilde{\beta}=a_0=0.25$, $m\alpha=1.5$ and $\xi=0$.}
	\label{fig04}
\end{figure}

We can note that in the case of $L/\tau \gg 1$ by using \eqref{formulaF},
the axial current goes to zero, as expected, with the fourth power of $\tau/L$.
The behavior of the axial current density is shown in Fig. \ref{fig04}
as a function of $r/\tau$ (left plot) and as a function of the compactification length $L/\tau$ (right plot) considering a minimal coulpling. Comparing the left plot with the result in the fermionic field case studied in \cite {mohammadi}, it is easy to see that they have opposite signs. This is not characteristic of the dS spacetime due to the fact that this also happens in the Minkowski spacetime studied in \cite{Eduardo} and \cite{Saha}. In both references the sign of the final axial currents should be corrected to the opposit one. The right plot in Fig. \ref{fig04} shows the  behavior of $\left\langle j^{z}(x)\right\rangle_{c}$ as a function of the compactification length $L/\tau$ for several values of $a_0$. As one would expect, it vanishes in the limit $L/\tau\to \infty$, i.e. in the absence of compactification.

\section{Conclusion}
\label{conclusion}
In this paper we have studied the VEVs of the induced bosonic currents in dS spacetime considering the presence of a compactified cosmic string.
These currents are induced by the presence of magnetic fluxes,
one along the string's core and another one enclosed by the compactified dimension. In order to perform our analysis, we constructed the positive frequency Wightman function associated with a massive scalar field considering that the field obeys the quasiperiodicity condition \eqref{eq03}, with a constant arbitrary phase $\beta$, and that the dimension along the $z$-direction is compactified to a circle with lenght $L$. The quasiperiodicity condition imposes that the quantum number related with the compactified string becomes discrete,  and in order to consider the summation
over this quantum number, we have employed the Abel-Plana summation formula \eqref{sumform}. By using this formula, the Wightman function and consequently the induced currents, are decomposed into two contributions: one induced by a cosmic string in dS spacetime with no compactification, Eq. \eqref{wightcs1}, and another one induced by the compactification, Eq. \eqref{wight2}. Taking into account these functions, we have calculate the azimuthal and axial current densities.

The string part of the azimuthal current density was obtained by substituting the Wightman function \eqref{wightcs1} into \eqref{eq21}. Its final expression is given by \eqref{jphicsFinal} and is an odd function of the magnetic flux along the string's core. In addition, the azimuthal current depends on the ratio $r/\tau$, which is the proper distance from the string measured in units of the dS curvature scale $\alpha$. Some particular asymptotic cases of \eqref{jphicsFinal} are considered.
For the region $r/\tau \gg 1$ the leading term of the azimuthal current induced by an uncompactified string, considering $\nu$ imaginary, is given by \eqref{jphicsLargeR}. From this expression we note that for large distances from the string, the azimuthal current density besides decaying with the fifth power in $\tau/r$,
has an oscillatory behavior if the parameter $\nu$ is imaginary. The parameter $\nu$ is
related to the mass of the scalar field. If $\nu$ is real, we have the same equation, but with the coefficient $1/2$. However, depending on the mass of the bosonic field, the decay behavior for large distances can be very different.
Considering a minimal coulpling, $\xi=0$, as mass tends to zero the string contribution of the azimuthal current density goes to zero with the second power of $\tau/r$. This is in contrast with the result for the fermionic field in the same geometry. In the fermionic field case the current decays with the fourth power of $\tau/r$ independent of the mass, and while decaying oscillates. However, in the massless limit the oscillation disappears. In the case of a conformal coupling and a massless scalar field, the azimuthal current density goes to zero with the fourth power of $\tau/r$, the same as for the fermionic field. These results for the bosonic field are
shown in Fig. \ref{fig01} where we have shown the string contribution of the azimuthal current density as a function of $r/\tau$ considering
just the case of a minimal coupling.

For the azimuthal current density induced by the compactification, we have used the Wightman function \eqref{wight2} and obtained the Eq. \eqref{jphicompFinal}. This expression is an odd function of the magnetic flux along the string
and an even function of the magnetic flux enclosed by the string axis. For regions close to the string, $r/\tau \ll 1$, and
considering that the parameter $\nu$ is imaginary, the leading term of the azimuthal
current density induced by the compactification is given by \eqref{jCompSmallR}. This expression presents a divergence on the string if $|a_0|\leq 1/q$ and is finite on the string if $|a_0|> 1/q$, in contrast with the fermionic case where the threshold value is $|a_0|=(1-1/q)/2$. We have also considered the case of $L/\tau \gg 1$, for $r/\tau$ fixed, where the leading contribution is given by \eqref{jphicLargL}.
In this regime, the azimuthal current density goes to zero, as one expects, with the fifth power of $\tau/L$. The behavior of the azimuthal current density induced by the compactification as functions of $r/\tau$ and the magnetic flux enclosed by the
string axis has been shown in the Fig. \ref{fig03} for a minimal coupling, $\xi=0$. The expression for the total azimuthal current density is presented in \eqref{jphitotal}. 

Our next step was to consider the axial current density. We have shown that the only non-zero contribution for the axial current is due to the compactification. The total axial current density is given by \eqref{jzc2}. We were able to decompose it into
two contributions. The first one is given by \eqref{jzk0} which is a purely topological contribution being dependent only on the compactification along the string axis. The other contribution for the azimuthal current density, given by \eqref{jzc3}, is
induced by the planar angle deficit, the magnetic flux and also the compactification. This contribution is an odd function of the magnetic flux enclosed by the compactified string and an even function of the magnetic flux running along
the string's core. Besides that, this contribution is finite on the string. We have also considered some particular cases in the axial current density expression. In Fig. \ref{fig04} we have plotted the axial current induced by the compactification as functions of $r/\tau$ and  $L/\tau$ considering a minimal coupling, $\xi=0$, for several values the parameter $q$ and
the magnetic flux along the string's core. We have shown that the induced axial current densities considering bosonic and fermionic fields have opposite signs which is a shared characteristic with the Minkowski spacetime.
Moreover, in the limit where $L/\tau \gg1$, both the azimuthal and axial current densities induced by the compactification go to zero, as one expects. Furthermore, we have shown that for a massless field and considering conformal coupling, $\xi=1/6$, the expressions for both current densities match the Minkowski one with the same geometry, found in \cite{Eduardo}, times the fourth power of the conformal factor $\tau/\alpha$. 

Before finishing this paper we would like to mention that all expressions obtained for the current densities associated with the uncompactified contribution present the natural dependence on the distance to the string, besides an explicit time dependence. In fact these dependencies appear in the combination $\tau/r$. As to the compactified contribution, an extra dependence on the compactification length $L$ and conformal time appears in the combination $\tau/L$. Finally we should say that our results may provide relevant information about the vacuum bosonic current in the inflationary phase of the Universe.

\section*{Acknowledgments}
E.A.F. B and A.M. would like to thank the Brazilian agency CAPES for the financial support. A.M. also thanks the Brazilian agency CNPq and
 Universidade Federal de Pernambuco Edital Qualis A for financial support.
%
\appendix

\section{Integral representation for the compactified Wightman function}
\label{compacified}
Here, we develop the intermediate steps in the calculation of the compactified Wightman function \eqref{wight2}. 
In order to obtain this function, we substitute \eqref{gf} into \eqref{sumform}, and also substitute the result into \eqref{eq18}. From the second term on the right hand side of \eqref{sumform}, the compactified Wightman function is given by
\begin{eqnarray}
\label{wcom}
W_c(x,x')&=&\frac{2iq(\tau\tau')^{3/2}e^{-ieA_z\Delta z}}{(2\pi)^3\alpha^2} \sum_{n=-\infty}^\infty e^{iqn\Delta\phi}
\int_0^\infty dp p J_{q|n+a|}(pr)J_{q|n+a|}(pr')\nonumber\\
&&\times \int_0^\infty dk_z\left[K_\nu(i\tau'\sqrt{(ik_z)^2+p^2})K_\nu(-i\tau\sqrt{(ik_z)^2+p^2})\right.\nonumber\\
&&\left.-K_\nu(i\tau'\sqrt{(-ik_z)^2+p^2})
K_\nu(-i\tau\sqrt{(-ik_z)^2+p^2})\right]\sum_{\lambda=\pm 1}\frac{e^{-\lambda k_z\Delta z}}{e^{Lk_z+2\pi i\lambda{\tilde{\beta}}}-1},
\nonumber\\
\end{eqnarray}
defining $u\equiv \frac{L}{2\pi}k_z$.
The integral over $k_z$ must be considered in two different situations as follows

\begin{itemize}
	\item For $p>k_z$:
	\begin{eqnarray}
	\label{ident1}
	\sqrt{(\pm ik_z)^2+p^2}=\sqrt{p^2-(k_z)^2}  \  .
	\end{eqnarray}
	\item For $p<k_z$:
	\begin{eqnarray}
	\label{ident2}
	\sqrt{(\pm ik_z)^2+p^2}=e^{\pm i\frac{\pi}{2}}\sqrt{(k_z)^2-p^2}  \  .
	\end{eqnarray}
\end{itemize}
The integral over the segment $[0, \ p]$ vanishes in contrast with the segment $[p, \ \infty)$. In fact in the latter interval the integrand reads
\begin{eqnarray}
\label{ident3}
K_\nu(e^{i\pi}\tau'\sqrt{(k_z)^2-p^2})K_\nu(\tau\sqrt{(k_z)^2-p^2})
-K_\nu(\tau'\sqrt{(k_z)^2-p^2}) K_\nu(e^{-i\pi}\tau\sqrt{(k_z)^2-p^2}) \ .
\end{eqnarray}
Now using the the identity \cite{Abra},
\begin{eqnarray}
K_\nu(z e^{im\pi})=e^{-im\nu\pi}K_\nu(z)-i\pi\sin(m\nu\pi)\csc(\nu\pi)I_\nu(z) \  , 
\end{eqnarray}
for $m$ integer, using the well-known relation involving the modified Bessel functions, $K_\nu(z)=\frac\pi 2 (I_{-\nu}(z)-I_\nu(z))/\sin{\nu\pi}$ as well as defining a new variable $\gamma^2=k_z^2-p^2$, Eq. \eqref{wcom} can be expressed as
\begin{eqnarray}
\label{wcom1}
W_c(x,x')&=&\frac{2q(\tau\tau')^{3/2}e^{-ieA_z\Delta z}}{(2\pi\alpha)^2}\sum_{n=-\infty}^\infty e^{iqn\Delta\phi}
\int_0^\infty dp \, p J_{q|n+a|}(pr)J_{q|n+a|}(pr')\nonumber\\
&&\times\int_0^\infty \frac{d\gamma \, \gamma}{\sqrt{\gamma^2+p^2}}\left[K_\nu(\tau'\gamma) I_{-\nu}(\tau\gamma)
+I_\nu(\tau'\gamma)
K_\nu(\tau\gamma)\right]\nonumber\\
&&\times\sum_{l=1}^\infty e^{-lL\sqrt{\gamma^2+p^2}}\cos(2\pi l{\tilde{\beta}}-i\Delta z\sqrt{\gamma^2+p^2}),
\end{eqnarray}
where we have also used the expansion $(e^{u}-1)^{-1}=\sum_{l=1}^\infty e^{-lu}$.

\begingroup\raggedright\endgroup

\end{document}